\documentstyle[amssymb,prl,aps,epsfig]{revtex}

\begin{document}
\title{Isotope effects in underdoped cuprate superconductors: a quantum critical
phenomenon}
\author{T. Schneider and H.Keller}
\address{Physik-Institut der Universit\"{a}t Z\"{u}rich, Winterthurerstr. 190,\\
CH-8057 Z\"{u}rich, Switzerland}
\date{\today}
\maketitle

\begin{abstract}
We show that the unusual doping dependence of the isotope effects on
transition temperature and zero temperature in - plane penetration depth
naturally follows from the doping driven 3D-2D crossover, the 2D quantum
superconductor to insulator transition (QSI) in the underdoped limit and the
change of the relative doping concentration upon isotope substitution. Close
to the QSI transition both, \ the isotope coefficient of transition
temperature and penetration depth approach the coefficient of the relative
dopant concentration, and its divergence sets the scale. These predictions
are fully consistent with the experimental data and imply that close to the
underdoped limit the unusual isotope effect on transition temperature and
penetration depth uncovers critical phenomena associated with the quantum
superconductor to insulator transition in two dimensions.

\end{abstract}


The observation of an unusual isotope effect in underdoped cuprate
superconductors on transition temperature \cite
{franck,schneikellerprl,schneikeller} and zero temperature penetration
depth\ \cite{zhaoprb,zhaon,zhao,hoferprl} \ poses a challenge to the
understanding of high temperature superconductivity. In this letter we show
that this behavior naturally follows from the doping driven 3D-2D crossover,
the 2D superconductor to insulator (QSI) transition in the underdoped limit
and the change of the relative doping concentration upon isotope
substitution. Thus, the unusual isotope effect on superconducting properties
of underdoped cuprates is derived to be a quantum critical phenomenon driven
by variation of the dopant concentration.

Consider the empirical phase diagram of La$_{2-x}$Sr$_{x}$CuO$_{4}$ \cite
{takagi,torrance,suzuki,kimura,yamada,nagano,fukuzumi,sasagawa,hoferdis}%
depicted in Fig.\ref{Fig:1}. It shows that after passing the so called
underdoped limit $\left( x=x_{u}\approx 0.05\right) $, where the quantum
insulator to superconductor (QSI) transition occurs \cite{book,klosters}, $%
T_{c}$ rises and reaches a maximum value $T_{c}^{m}$ at $x_{m}\approx 0.15$.
With further increase of $x$, $T_{c}$ decreases and finally vanishes in the
overdoped limit $x_{o}\approx 0.3$. This phase transition line $T_{c}\left(
x\right) $, separating the superconducting from the normal conducting phase,
appears to be a generic property of cuprate superconductors. In $\ $La$_{2-x}
$Sr$_{x}$CuO$_{4}$, HgBa$_{2}$CuO$_{4+x}$ \cite{fukuokahg,hoferhg} and Bi$%
_{2}$Sr$_{2}$CuO$_{6+x}$ \cite{groenbi}both, the underdoped and overdoped
limits, corresponding to critical endpoints, are experimentally accessible.
In other cuprates, including Bi$_{2}$Sr$_{2}$CaCu$_{2}$O$_{8+x}$ \cite
{groenbi}, YBa$_{2}$Cu$_{3}$O$_{7-x}$ \cite{chien} and Y$_{1-x}\Pr_{x}$Ba$%
_{2}$Cu$_{3}$O$_{7-\delta }$ \cite{neumeierpr}, only the underdoped and
optimally doped regimes appear to be accessible. As shown in Fig.\ref{Fig:1}
for La$_{2-x}$Sr$_{x}$CuO$_{4}$ the effective mass anisotropy, measured in
terms of $\gamma =\sqrt{M_{c}/M_{ab}}$ increases drastically by approaching
the underdoped limit. This property, also observed in HgBa$_{2}$CuO$%
_{4+\delta }$ \cite{hoferhg} and YBa$_{2}$Cu$_{3}$O$_{7-x}$ \cite{chien},
appears to be generic and reveals the crossover from three (3D) to two
dimensional (2D) behavior.

In our considerations the starting point is the critical endpoint of the
phase transition line $T_{c}\left( x\right) $, where at $T=0$ and $x=x_{u}$,
driven by variation of the dopant concentration $x$, \ the QSI transition
occurs. Close to this critical endpoint the bulk superconductors correspond
to a stack of independent superconducting slabs of thickness $d_{s}$.
Moreover, close to such a critical point, low energy properties depend only
on the spatial dimensionality of the system, the number of components of the
order parameter and the range of the interaction. The theory of quantum
critical phenomena predicts that the transition temperature and zero
temperature in - plane penetration depth scale as \cite{kim,book,klosters}

\begin{equation}
T_{c}=a\ \delta ^{z\overline{\nu }},\ \ \ \ \ \ \frac{1}{\lambda
_{ab}^{2}\left( 0\right) }=b\delta ^{z\overline{\nu }}.  \label{eq1}
\end{equation}
$z$ is the dynamic critical exponent, $\overline{\nu }$ the exponent of the
diverging length, $a$ and $b$ nonuniversal critical amplitudes, and 
\begin{equation}
\delta =x-x_{u}  \label{eq2}
\end{equation}
measures the distance from the quantum critical point at $x_{u}$, where in
cuprate superconductors the QSI transition occurs. For a complex scalar
order parameter and in D=2 the critical amplitudes $a$ and $b$ and the slab
thickness $d_{s}$ are not independent but related by the universal relation 
\cite{kim,book,klosters} 
\begin{equation}
\mathrel{\mathop{\lim }\limits_{\delta \rightarrow 0}}%
\frac{T{_{c}\lambda _{ab}^{2}(0)}}{d_{s}}=\frac{a}{bd_{s}}=\frac{1}{%
\overline{Q}_{2}}\left( {\frac{\Phi _{0}^{2}}{16\pi ^{3}k_{B}}}\right) ,
\label{eq3}
\end{equation}
where $\overline{Q}_{2}$ is a universal number. Although the experimental
data are rather sparse close to the QSI transition, the overall picture
turns out to be highly suggestive and provides consistent evidence for the
QSI transition in D=2 at the critical endpoint in the underdoped limit with
critical exponents $z\approx 1$ and $\overline{\nu }\approx 1$\cite
{book,klosters}. This estimate is close to theoretical predictions \cite
{fisher,herbut}, from which $z=1$ is expected for a bosonic system with
long-range Coulomb interactions independent of dimensionality and $\overline{%
\nu }\geq 1\approx 1.03$ in $D=2$. \ In \ Fig.\ref{Fig:1} it is seen that $%
\overline{\nu }\approx 1$ is also consistent with the doping dependence of
the effective mass anisotropy $\gamma =\sqrt{M_{c}/M_{ab}}\propto \delta ^{-%
\overline{\nu }}$.

We are now prepared to explore the implications of the QSI transition on the
isotope effect. From the definition of the isotope coefficient 
\begin{equation}
\beta _{T_{c}}=-\frac{m}{T_{c}}\frac{dT_{c}}{dm}  \label{eq4}
\end{equation}
and Eq. (\ref{eq1}) we obtain the scaling expression 
\begin{equation}
\beta _{T_{c}}=\beta _{a}+\beta _{\delta },  \label{eq5}
\end{equation}
where 
\begin{equation}
\beta _{a}=-\frac{m}{a}\frac{da}{dm},\ \ \beta _{\delta }=\frac{z\overline{%
\nu }}{\delta }\overline{\beta }_{\delta },\ \ \overline{\beta }_{\delta }=-m%
\frac{d\delta }{dm}.  \label{eq6}
\end{equation}
Since in the doping regime of interest, isotope substitution lowers the
transition temperature \cite{franck,schneikeller}, while the dopant
concentration $x$ remains nearly unchanged \cite{zhao}, there is a positive
shift of the underdoped limit $x_{u}$, and $\overline{\beta }_{\delta }$
reduces to 
\begin{equation}
\overline{\beta }_{\delta }=-m\frac{d\delta }{dm}\approx m\frac{dx_{u}}{dm}%
>0.  \label{eq7}
\end{equation}
Thus, by approaching the QSI transition $\beta _{\delta }$ diverges as $%
\beta _{\delta }\propto \delta ^{-1}$ [Eq.(\ref{eq4})] and provided that $%
\beta _{a}$ remains finite, $\beta _{T_{c}}$ is predicted to tend to $\beta
_{\delta }$ so that 
\begin{equation}
\frac{1}{\beta _{T_{c}}}\rightarrow \frac{1}{\beta _{\delta }}=\overline{r}\
\left( \frac{T_{c}}{T_{c}^{m}}\right) ^{1/z\overline{\nu }},\ \ \ \ \ 
\overline{r}=\frac{T_{c}^{m}}{z\overline{\nu }\overline{\beta }_{\delta
}a^{1/z\overline{\nu }}}.  \label{eq8}
\end{equation}
Here we expressed $\delta $ in terms of $T_{c}$ [Eq.(\ref{eq1})] and
rescaled $T_{c}$ by $T_{c}^{m}$, the transition temperature at optimum
doping, to reduce variations of $T_{c}$ between different cuprates \cite
{schneikellerprl,schneikeller}.

To verify this prediction we show $1/\beta _{_{T_{c}}}$ versus $%
T_{c}/T_{c}^{m}$ for La$_{1.85}$Sr$_{0.15}$Cu$_{1-x}$O$_{4}$ \cite
{babushkina}, YBa$_{2-x}$La$_{x}$Cu$_{3}$O$_{7}$ \cite{bornemann} and Y$%
_{1-x}\Pr_{x}$Ba$_{2}$Cu$_{3}$O$_{7}$ \cite{francky} in Fig.\ref{Fig:2}. As
predicted, by approaching the underdoped limit, corresponding to $%
T_{c}/T_{c}^{m}=0$, the data collapses on a single curve, consistent with a
straight line pointing to the expected value $z\overline{\nu }\approx 1$ ,
yielding the estimate $\overline{r}\approx 6$ and confirming that $\beta
_{a} $ is finite and $\overline{\beta }_{\delta }>0$. \ Thus, the OSI
transition in $D=2$ accounts naturally for the unusual doping dependence of $%
\beta _{T_{c}}$ and identifies the isotope induced change of \ the
underdoped limit as the main mechanism. Notice that the empirical estimate
for $\overline{r}$ implies that the maximum transition temperature is fixed
by $T_{c}^{m}\approx 6a\ \overline{\beta }_{\delta }$.

Another important element brought by the QSI transition in D=2 \ is the
universal relation given in Eq. (\ref{eq3}). It implies with Eq.(\ref{eq5})
that the isotope coefficients of \ $T_{c}$, $1/\lambda _{ab}^{2}$ , critical
amplitudes $a$ and $b$ and slab thickness $d_{s}$ are related by 
\begin{equation}
\beta _{T_{c}}=\beta _{1/\lambda _{ab}^{2}}+\beta _{d_{s}}=\beta _{a}+\beta
_{\delta },\ \ \ \ \beta _{a}=\beta _{b}+\beta _{d_{s}},  \label{eq9}
\end{equation}
where $\beta _{F}=-\frac{m}{F}\frac{dF}{dm}$ and $F=T_{c},$ $1/\lambda
_{ab}^{2},\ d_{s},\ a$ and $b$. Noticing that $\beta _{a}$ was confirmed to
be bounded, this is also true for $\beta _{b}$ and $\beta _{d_{s}}$. Since $%
\beta _{T_{c}}$ diverges as $\beta _{T_{c}}=\beta _{\delta }=\left(
T_{c}/T_{c}^{m}\right) ^{-1/z\overline{\nu }}/\overline{r}$ [Eq.(\ref{eq8}], 
$\beta _{1/\lambda _{ab}^{2}}$ is predicted to approach 
\begin{equation}
\beta _{1/\lambda _{ab}^{2}}=\beta _{T_{c}}=\beta _{\delta }=\frac{1}{%
\overline{r}}\left( \frac{T_{c}}{T_{c}^{m}}\right) ^{-1/z\overline{\nu }}
\label{eq10}
\end{equation}
close to the QSI transition. Although the experimental data for $\beta
_{1/\lambda _{ab}^{2}}$ and $\beta _{T_{c}}$ on identical samples are rather
sparse, the results shown in Fig.\ref{Fig:3} for the oxygen isotope effect
in La$_{2-x}$Sr$_{x}$Cu$_{1-x}$O$_{4}$ \cite{zhao,hoferprl} clearly confirm
this prediction. Indeed as the underdoped limit is approached the data
points tend to the solid line, marking $1/\beta _{T_{c}}=1/\beta _{1/\lambda
_{ab}^{2}}$.

In conclusion, we have shown that the strong doping dependence of the
isotope effects on transition temperature and zero temperature in - plane
penetration depth naturally follows from the doping driven 3D-2D crossover
and the 2D quantum superconductor to insulator transition in the underdoped
limit. As the quantum superconductor to insulator transition is approached,
the isotope coefficient of transition temperature $\beta _{T_{c}}$ and
penetration depth $\beta _{1/\lambda _{ab}^{2}}$ tend to the coefficient of
the relative dopant concentration $\beta _{\delta }$, and its divergence
sets the scale. These predictions are fully consistent with the experimental
data and imply that close to the underdoped limit the unusual isotope
effects on transition temperature and penetration depth uncovers critical
phenomena, associated with the quantum superconductor to insulator
transition in two dimensions. Moreover, since lattice distortions are the
primary consequence of isotope substitution it becomes clear that the
isotope effects on transition temperature and zero temperature penetration
depth probe the effect of lattice degrees of freedom on the condensate.
However, until now most of the proposed theories of high temperature
superconductivity in the cuprates have considered charge and spin
fluctuations in a doped two dimensional (2D) antiferromagnet and disregarded
any coupling to the lattice degrees of freedom.

\bigskip

The authors are grateful to K.A. M\"{u}ller, S. Roos and G.M. Zhao for very
useful comments and suggestions on the subject matter. This work was
partially supported by the Swiss National Science Foundation.

\begin{figure}[tbp]
\caption{$\ T_{c}$ and $\protect\gamma $ versus $x$ for La$_{2-x}$ Sr$_{x}$%
CuO$_{4}$. $T_{c}$ data taken from [8-16 ]. The solid curve corresponds to $%
\protect\gamma \propto \protect\delta ^{-\overline{\protect\nu }}$ with $%
\overline{\protect\nu }=1$, $\protect\delta =x-x_{u}$ and $x_{u}=0.05$.}
\label{Fig:1}
\end{figure}

\begin{figure}[tbp]
\caption{Inverse isotope coefficient $1/\protect\beta _{T_{c}}$ versus $%
T_{c}/T_{c}^{m}$ for La$_{1.85}$Sr$_{0.15}$Cu$_{1-x}$O$_{4}$ [27], YBa$%
_{2-x} $La$_{x}$Cu$_{3}$O$_{7}$ [28]\ and Y$_{1-x}\Pr_{x}$Ba$_{2}$Cu$_{3}$O$%
_{7}$ [29]. The straight line corresponds to Eq.(8) with $z\overline{\protect%
\nu }=1$ and $\overline{ r}=6$.}
\label{Fig:2}
\end{figure}

\begin{figure}[tbp]
\caption{Oxygen isotope effect for underdoped La$_{2-x}$Sr$_{x}$CuO$_{4}$
[6,7] in terms of $\ -T_{c}/\Delta T_{c}\propto 1/\protect\beta _{T_{c}}$ \
versus $\ -\left( 1/{\protect\lambda _{ab}^{2}(0)}\right) /\Delta \left( 1/{%
\protect\lambda _{ab}^{2}(0)}\right) \propto \protect\beta _{1/\protect%
\lambda _{ab}^{2}}$. The solid line corresponds to $\protect\beta _{d_{s}}=0$%
\ in Eq.(9) and the dashed one is a guide to the eyes. {\protect\Large $%
\bullet $} : taken from [6] and $\blacksquare :$ taken from [7].}
\label{Fig:3}
\end{figure}

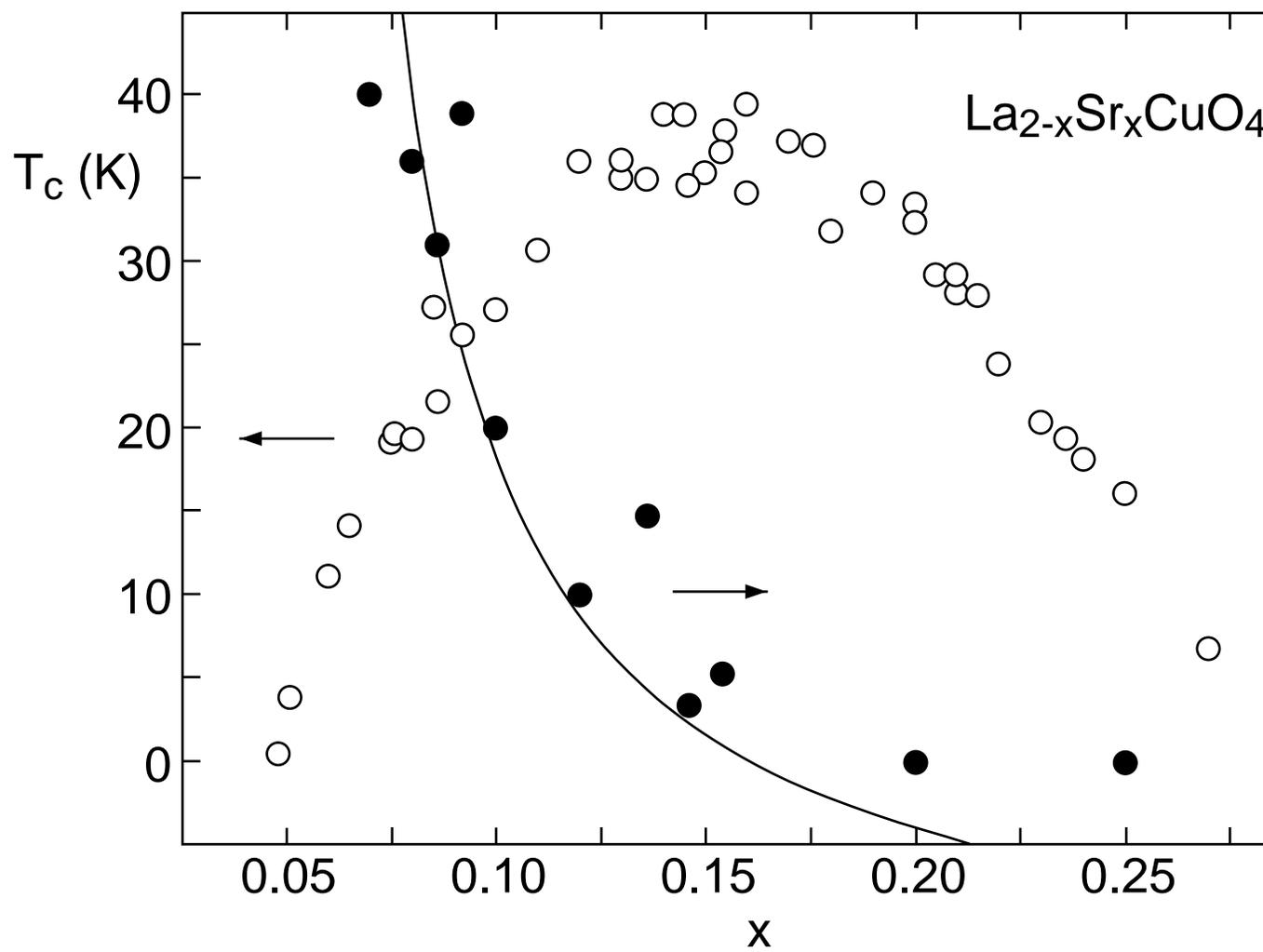
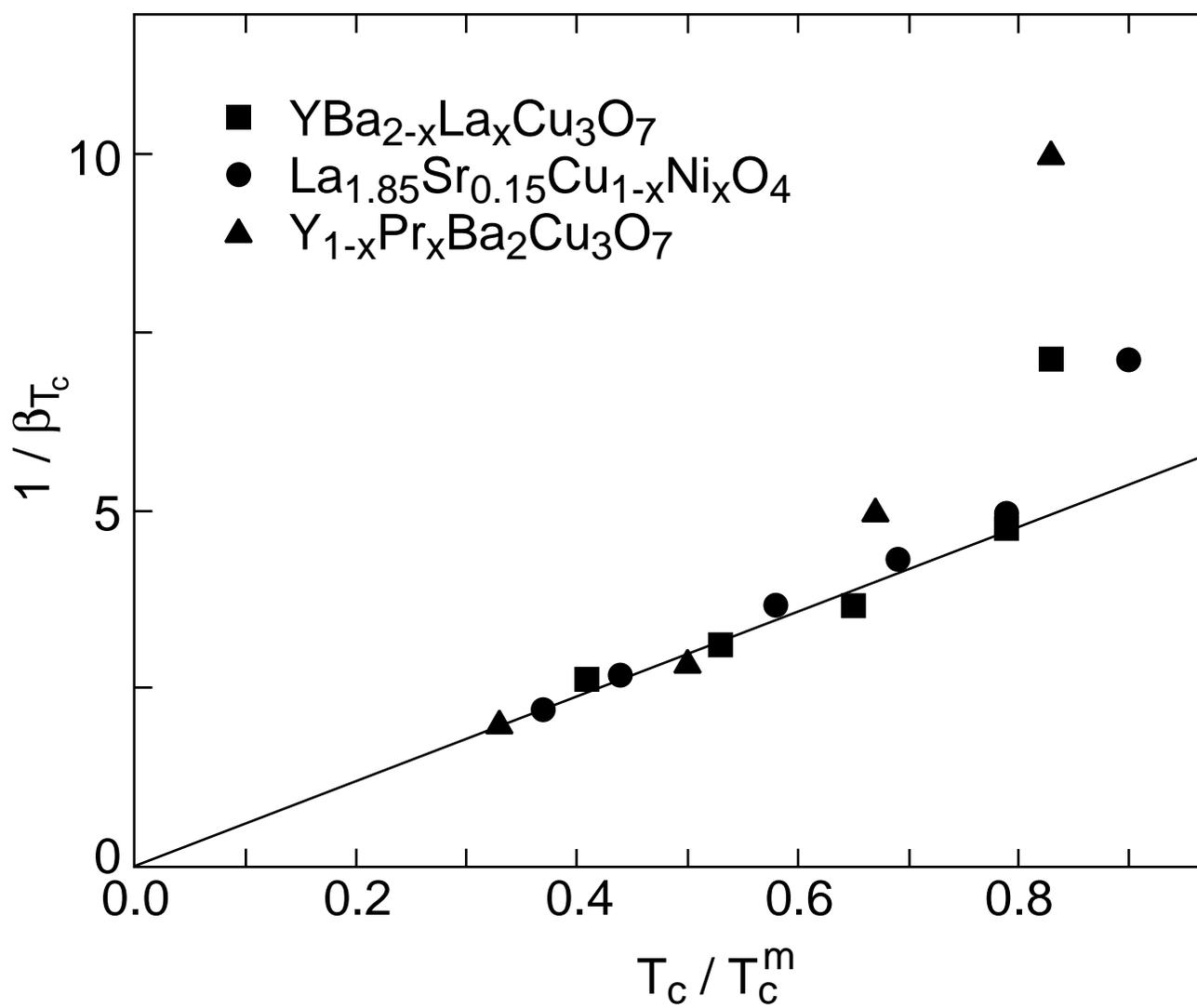
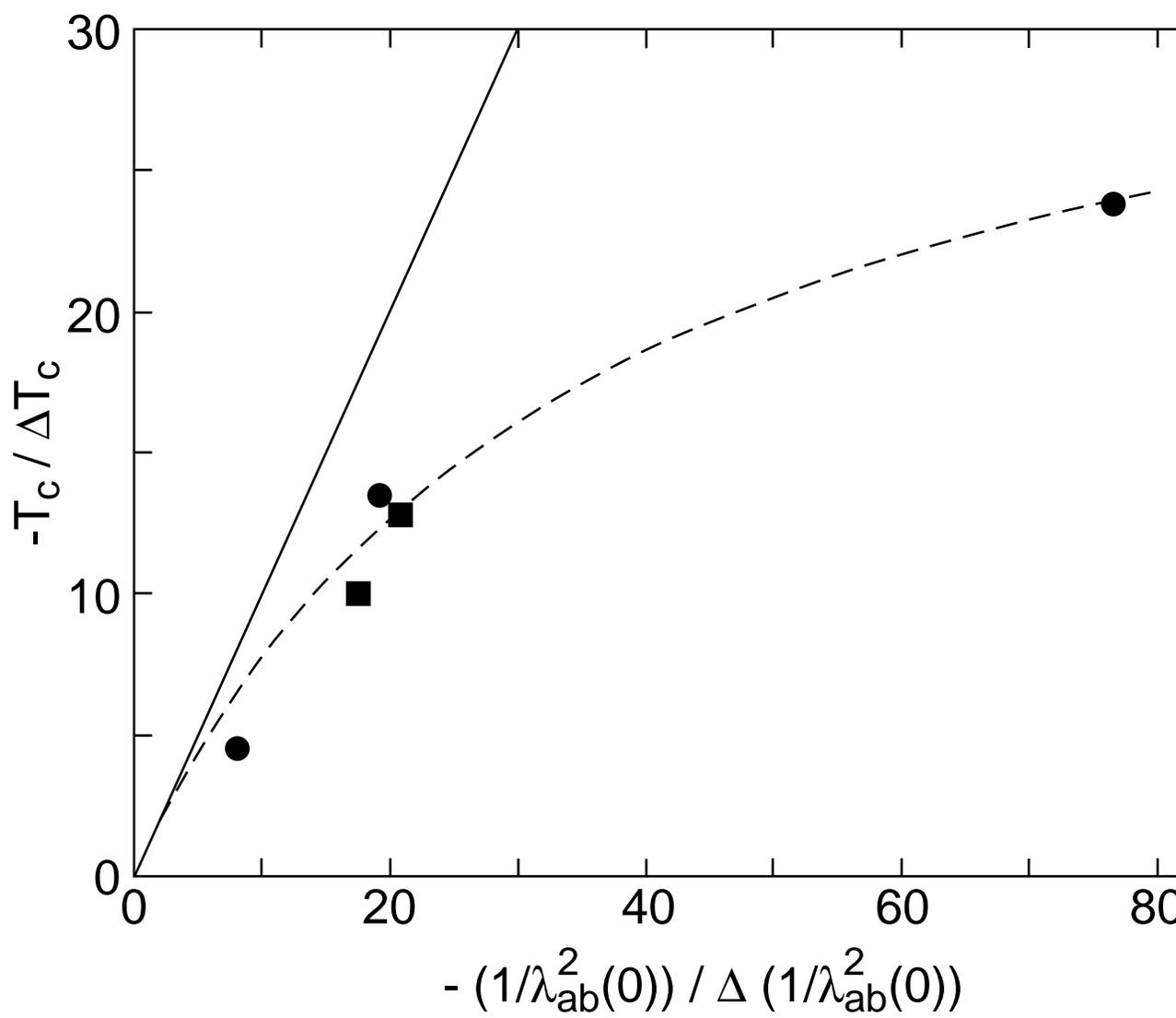
\end{document}